\documentclass[twocolumn,showpacs,aps,prd,superscriptaddress,floatfix]{revtex4}
\usepackage{graphicx}
\usepackage{dcolumn}
\usepackage{amsmath}
\usepackage{epsfig}

\newcommand{\BaBarType}      {PUB}  
\newcommand{\BaBarYear}       {07}
\newcommand{\BaBarNumber}     {030}
\newcommand{\SLACPubNumber} {12694}

\def\emm{\overrightarrow{m}}
\def\ix{\overrightarrow{x}}
\def\dbarp{\tilde{D}^0}

\def\Dstarz  {\ensuremath{D^{*0}}\xspace}
\def\Dstarm  {\ensuremath{D^{*-}}\xspace}

\def\Kstarz  {\ensuremath{K^{*0}}\xspace}

\def\Kstarp  {\ensuremath{K^{*+}}\xspace}

\def\Kpm      {\ensuremath{K^{\pm}}\xspace}
\def\Dz      {\ensuremath{D^0}\xspace}
\def\piz   {\ensuremath{\pi^{0}}\xspace}
\def\pim   {\ensuremath{\pi^{-}}\xspace}
\def\pip   {\ensuremath{\pi^{+}}\xspace}
\def\KS    {\ensuremath{K_{\scriptscriptstyle S}}\xspace}

\def\mes        {\mbox{$m_{\rm ES}$}\xspace}

\def\to         {\ensuremath{\rightarrow}\xspace}

\def\Dzbpar  {\ensuremath{\Dbar^{0}}\xspace}
\def\rads  {\ensuremath{R_{ADS}}}
\def\rblim {0.19}
\def\rbmeas{\ensuremath{0.091\pm0.059}}
\def\radsval{0.012}
\def\radsstatp{+0.012}
\def\radssysp{+0.010}
\def\radsstatm{-0.010}
\def\radssysm{-0.007}
\def\radssys{0.0076}
\def\radsmeas{\ensuremath{\radsval^{\radsstatp}_{\radsstatm}({\rm 
stat})^{\radssysp}_{\radssysm}({\rm sys}})}
\def\radslim{0.039}
\input pubboard/babarsym

\long\def\inst#1{\par\nobreak\kern 4pt\nobreak
    {\it #1}\par\vskip 10pt plus 3pt minus 3pt}

\begin{document}
{\pagestyle{empty}

\begin{flushleft}
\babar-\BaBarType-\BaBarYear/\BaBarNumber \\ 
SLAC-PUB-\SLACPubNumber 
\end{flushleft}

\title{Search for {\boldmath $b \rightarrow u$} transitions in
{\boldmath $B^- \to [\Kp\pim\piz]_D K^-$}}

\begin{abstract}
\noindent
We search for decays of a $B$ meson into a neutral $D$
 meson and a charged kaon,
with the $D$ meson decaying into a charged kaon, a charged pion, and a neutral pion.
This final state
can be reached through the $b \to c$
transition $B^- \to D^{0}K^-$ followed by the doubly-Cabibbo-suppressed 
$D^0 \to K^+ \pi^-\piz$, or the $b \to u$
transition $B^- \to \Dzb K^-$ followed by the Cabibbo-favored $\Dzb
\to K^+ \pi^-\piz$.
The interference of these two amplitudes
is sensitive to the angle $\gamma$ of the unitarity triangle.
We present results 
based on 226$\times 10^{6}$ $\epem\to\FourS \to B\Bbar$ events collected with the
\mbox{\slshape B\kern-0.1em{\smaller A}\kern-0.1em
    B\kern-0.1em{\smaller A\kern-0.2em R}}\ detector at SLAC. 
We find no significant evidence
for these decays and 
 we set a limit  $\rads\equiv
   \frac{\Gamma([K^+ \pi^-\piz]_D K^-)+\Gamma([K^- \pi^+\piz]_D K^+)}{\Gamma([K^+ \pi^-\piz]_D K^+)+\Gamma([K^- \pi^+\piz]_D K^-)}
 < \radslim$ at 95\% confidence level,
which we translate with a Bayesian approach into
$r_B \equiv |A(B^- \to \Dzb K^-)/A(B^- \to \Dz K^-)| < \rblim$ at  
95\% confidence level. 

\end{abstract}

\pacs{13.25.Hw, 14.40.Nd}
%
\author{B.~Aubert}
\author{M.~Bona}
\author{D.~Boutigny}
\author{Y.~Karyotakis}
\author{J.~P.~Lees}
\author{V.~Poireau}
\author{X.~Prudent}
\author{V.~Tisserand}
\author{A.~Zghiche}
\affiliation{Laboratoire de Physique des Particules, IN2P3/CNRS et Universit\'e de Savoie, F-74941 Annecy-Le-Vieux, France }
\author{J.~Garra~Tico}
\author{E.~Grauges}
\affiliation{Universitat de Barcelona, Facultat de Fisica, Departament ECM, E-08028 Barcelona, Spain }
% ...
\collaboration{The \babar\ Collaboration}
\noaffiliation

\maketitle

} 
\section{Introduction}
\begin{figure}[hb]
   \begin{center}
      \epsfig{file=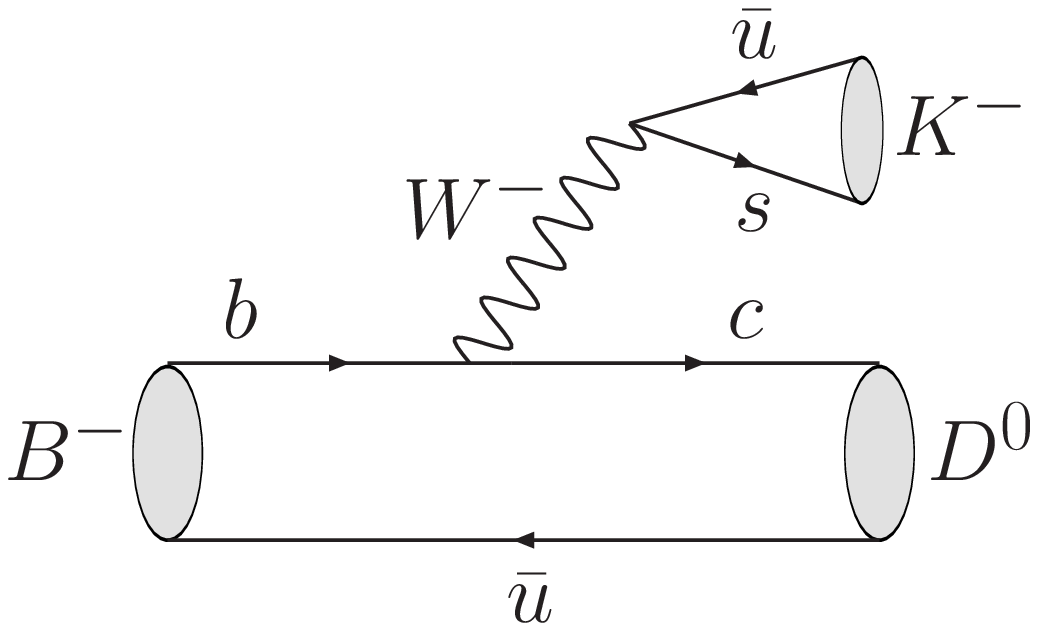,width=0.47\linewidth}
      \epsfig{file=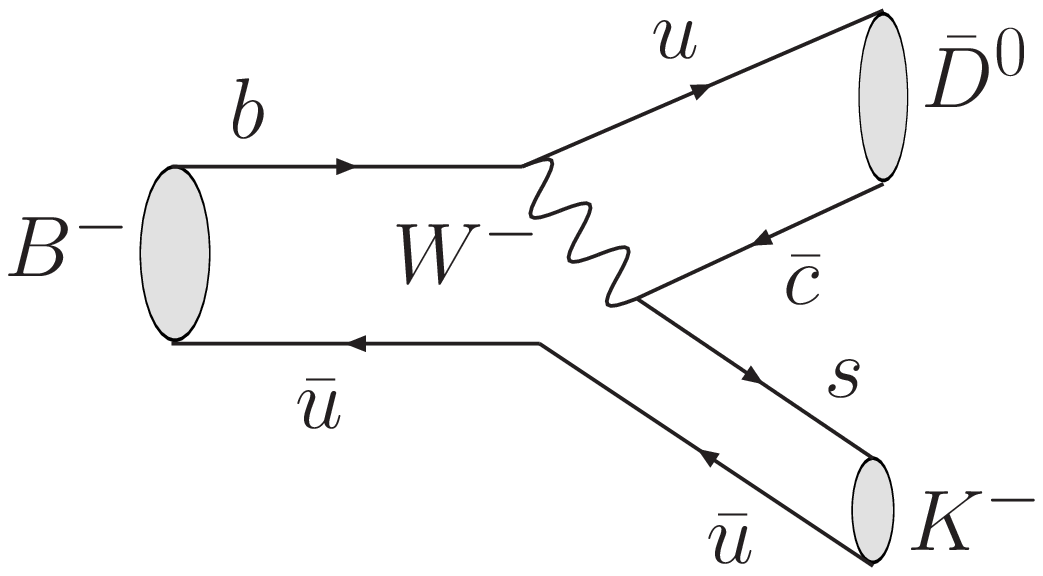,width=0.47\linewidth}
   \caption{Feynman diagrams for the CKM-favored
 $B^- \to D^{0} K^{-}$ and 
the CKM- and color-suppressed $B^-\to \Dzbpar K^{-}$ decays.}
   \label{fig:feynman}
   \end{center}
   \end{figure}

   Following the discovery of \CP violation in $B$ meson 
   decays and the measurement of the angle $\beta$
   of the unitarity triangle~\cite{cpv} associated with
   the Cabibbo-Kobayashi-Maskawa (CKM) quark mixing matrix, the focus has turned
   toward the measurements of the other angles $\alpha$ and $\gamma$.
   Following Ref.~\cite{dk1}, several  methods have been proposed to measure the relative weak phase
   between the $B^- \to D^{0}K^{-}$ amplitude, proportional to the CKM matrix element $V_{cb}$ (Fig.~\ref{fig:feynman}),
   and the $B^- \to \Dzbpar K^{-}$ amplitude, proportional to $V_{ub}$. 
This weak phase, which by definition is $\gamma={\rm arg}(-V_{ub}^*V^{}_{ud}/V_{cb}^*V^{}_{cd})$,
 can be measured from the interference that occurs when the $D^{0}$ and the
   \Dzb\ decay to common final states.

   As an extension of the method proposed in Ref.~\cite{dk2}, we search for 
   $B^- \to [K^+\pi^-\piz]_D K^-$~\cite{cc},
   where the CKM-favored  $B^- \to D^{0} K^-$ decay,
   followed by
   the doubly Cabibbo-suppressed $\Dz \to K^+ \pi^-\piz$ decay,
   interferes with the CKM-suppressed
   $B^- \to \Db^{0} K^-$ decay,
   followed by the Cabibbo-favored $\Dzb \to K^+ \pi^-\piz$ decay.

   In order to reduce the systematic uncertainties, 
we measure ratios of decay rates:
   \begin{eqnarray}\label{eq:rads}
   \rads &\equiv& 
   \frac{\Gamma([K^+ \pi^-\piz]_D K^-)+\Gamma([K^- \pi^+\piz]_D K^+)}{\Gamma([K^+ \pi^-\piz]_D K^+)+\Gamma([K^- \pi^+\piz]_D K^-)}\nonumber
\\
   &=& r_B^{2} + r_{D}^{2} + 2 r_B r_{D} C \cos\gamma,\\\nonumber
{A}_{ADS} &\equiv&    \frac{\Gamma([K^+ \pi^-\piz]_D K^-)-\Gamma([K^- \pi^+\piz]_D K^+)}{\Gamma([K^+ \pi^-\piz]_D K^-)+\Gamma([K^- \pi^+\piz]_D K^+)}\\
&=& 2r_B r_{D}S\sin\gamma/\rads,\label{eq:aads}
   \end{eqnarray}
   \noindent  where $r_B \equiv \left| \frac{A(B^- \to \Dzb K^-)}{A(B^- \to D^0 K^-)} \right|$, 
$r^2_D \equiv  \frac{\BR(D^0 \to K^+ \pi^-\piz)}{\BR(D^0 \to K^- \pi^+\piz)} $.  The $C$ and $S$ 
parameters are defined as: 
\begin{eqnarray}\label{eq:cs}
C = \frac{\int \mathcal{A}_{D}(\emm) \overline{\mathcal{A}}_{D}(\emm) \cos(\bar{\delta}(\emm)-\delta(\emm)+\delta_B) d\emm}{\sqrt{\int|\overline{\mathcal{A}}_{D}(\emm)|^{2} d\emm \cdot \int|\mathcal{A}_{D}(\emm)|^{2} d\emm }},\label{eqC}
\\
S =  \frac{\int \mathcal{A}_{D}(\emm) \overline{\mathcal{A}}_{D}(\emm) \sin(\bar{\delta}(\emm)-\delta(\emm)+\delta_B) d\emm}{\sqrt{\int|\overline{\mathcal{A}}_{D}(\emm)|^{2} d\emm \cdot \int|\mathcal{A}_{D}(\emm)|^{2} d\emm }}\, ,
\end{eqnarray}
where $\emm$ indicates a point in the Dalitz plane 
$[m_{K\pi}^2, m_{K\piz}^2]$, $[{\mathcal{A}}_D(\emm),\delta(\emm)]$  ($[\overline{\mathcal{A}}_D(\emm),\bar{\delta}(\emm)]$) the absolute value and
the strong phase of the $\Dz$ (\Dzb) decay amplitude,
and $\delta_B$ the strong phase difference between the two interfering $B$ decay amplitudes.  Equations \ref{eq:rads} and \ref{eq:aads} hold when neglecting $D$-mixing effects, which in the Standard Model (SM) give 
negligible corrections to $\gamma$ \cite{Dmix} and do not affect the $r_B$ measurement.

Determining the angle $\gamma$ from the measurements of \rads\ and $A_{ADS}$
 requires extracting  the strong phases by means of a Dalitz analysis of the 
three-body decay of the neutral $D$ meson, for which the available statistics are insufficient. 
However, with the current statistics we can measure \rads\ and constrain $r_B$ 
by exploiting the fact that in Eq.~\ref{eq:rads} $|C|\leq 1$.
Since  the value of $r_B$ is related to the level of interference between the diagrams
of Fig.~\ref{fig:feynman}, high values of $r_B$ lead to a better sensitivity to $\gamma$
in any measurement involving $B\to\Dz K$ decays. Thus, $r_B$ is a  key
ingredient for the extraction of $\gamma$ from
other measurements~\cite{dalitz}.

Both the Belle and \babar\ collaborations have published similar measurements 
but in a different decay chain, $B\to D K$  with 
$D\to K\pi$~\cite{oldADS}. Unlike those measurements, we can take advantage
of the smaller value of  $r_D$, given by 
$r^2_D=(0.214\pm0.008\pm0.008)\%$~\cite{kpipiz} in 
$D\to K\pi\piz$ decays as opposed to $r^2_D=(0.362\pm 
0.020\pm0.027)\%$~\cite{rdkpi} in 
$D\to K\pi$ decays. This implies that for a given error on \rads, the sensitivity 
to $r_B$ is better. 

\section{Event Reconstruction and Selection}

   The results presented in this paper are based on 
    226$\times 10^{6}$ $\FourS\to\BB$
decays collected 
   between 1999 and 2004 with the \babar\ detector at the \pep2\
   \BF\ at SLAC~\cite{pep2}.  
   Approximately 7\% of the collected data (15.8~fb$^{-1}$) have a 
   center-of-mass (CM) energy 40~\mev below the $\FourS$ resonance. These ``off-resonance'' data are
   used to study backgrounds from continuum events, $e^+ e^- \to q \bar{q}$
   ($q=u,d,s,$ or $c$).
The \babar\ detector is described elsewhere~\cite{babar}. Charged-particle tracking is provided by a five-layer silicon
vertex tracker (SVT) and a 40-layer drift chamber (DCH).
In addition to providing precise position information for tracking, the SVT and DCH
also measure the specific ionization ($dE/dx$), which is used for particle
identification of low-momentum charged particles. At higher momenta ($p>0.7$~\gevc)
pions and kaons are identified by Cherenkov radiation detected in a ring-imaging
device (DIRC). 
The position and energy of photons are
measured with an electromagnetic calorimeter (EMC) consisting of 6580 thallium-doped CsI crystals.
These systems are mounted inside a 1.5T solenoidal super-conducting magnet.

   The event selection was developed from studies of off-resonance
   data and $B\Bbar$ and continuum events simulated with Monte Carlo (MC) techniques. A large on-resonance data sample of $B^- \to D^0 \pi^-$,
   $D^0 \to K^- \pi^+\piz$ events was used to validate several aspects of the
   simulation and analysis procedure. We refer to this mode
   as $B \to D\pi$.

   Both kaon candidates are required to satisfy kaon identification criteria, which are based on the 
specific ionization loss
measured in the tracking devices and on the Cherenkov angles measured in the DIRC
 and are typically 85\% efficient, depending on momentum and polar angle.
   Misidentification rates are at the 2\% level. The \piz\ candidates 
   are reconstructed as pairs of photon candidates in the EMC, each with energy larger than 70\mev
   and a lateral shower profile consistent with an electromagnetic deposit. These pairs must have a total
energy greater than 200\mev and $118 <m_{\gamma\gamma}<145 \mevcc$.  
   To account for the correlation between the tails in the distribution of the $K\pi\piz$ invariant mass and the \piz\ candidate mass,
we require the difference between the two measured masses to
 be within 32.5 \mevcc of the expected value of $m_{D^0}-m_{\pi^0} =$ 1729.5\mevcc~\cite{PDG}, retaining 90\% of the signal.
   The remaining background from other $B^\pm \to [h_1 h_2\piz]_D h_3^\pm$~\cite{cc} modes
   is reduced by removing events where the invariant mass of any $h_1 h_2\piz$
   candidate, with any particle-type assignment other than the signal hypothesis,
 is consistent with the $D^0$ meson mass, retaining 92\% of the signal.

   After these requirements, the background is mostly due to \Dz - \Dzb 
pair production in 
   $e^+ e^- \to c \bar{c}$ events, with 
   $\Dzb \to K^+ \pi^-\piz$ and $D \to K^-$.
   To discriminate between the signal and this dominant background we use a neural network ($NNet$) with
   six quantities that distinguish continuum and $B\Bbar$ events:  
    $L_0 = \sum_i{p_i}$ and  $L_2 = \sum_i{p_i \cos^2\theta_i}$, both
   calculated in the CM frame, where $p_i$ is the 
momentum of particle i, $\theta_i$ is its angle relative to the thrust axis 
of the $B$ candidate, and the sum runs over all tracks and clusters not 
used to reconstruct the $B$ candidate;
  the angle in
   the CM frame between the thrust axes of the $B$ candidate
   and of the detected remainder of the event;
   the polar angle of the $B$ candidate
   in the CM frame;
   the distance of closest approach between the 
   track of the kaon candidate from the $B$ and the trajectory of the 
reconstructed $D$ meson 
(this is
   consistent with zero for signal events, but can be
   larger in $c\bar{c}$ events);
   the distance along the beams between the reconstructed vertex
   of the $B$ candidate and the vertex of the other tracks in the event, this is consistent
   with zero for continuum events, but is sensitive to the $B$ lifetime for signal events.

   The {\it NNet} is trained with simulated continuum and signal events.  
   We find agreement between the distributions of all six variables
   in simulation, off-resonance data,
   and $B \to D \pi$ events.  We apply a loose pre-selection  on the {\it NNet} ($0.4<NNet<1.0$) with a 90\% efficiency 
   for signal and a 68\% rejection power for continuum, and then use the {\it NNet} itself in the likelihood
   fit described below to fully exploit its discriminating power.

   A $B$-meson candidate is characterized by the energy-substituted mass
   $\mes \equiv \sqrt{(\frac{s}{2}  + \vec{p}_0\cdot \vec{p}_B)^2/E_0^2 - p_B^2}$
   and energy difference $\Delta E \equiv E_B^*-\frac{\sqrt{s}}{2}$, 
   where $E$ and $p$ are energy and momentum, the asterisk
   denotes the CM frame, the subscripts $0$ and $B$ refer to the
   initial $\epem$ state and $B$ candidate, respectively, and $s$ is the square
   of the CM energy.  For signal events, $\mes$ is centered around the $B$ mass with a 
   resolution of about 2.5 \mevcc, and $\Delta E$ is centered at zero with
a resolution of 17 \mev.

   Considering both the case where the two kaons have 
the same and the opposite charge (referred to as "same-sign" and 
"opposite-sign" samples respectively), 28621 
events 
   survive the selection described above and the loose 
   requirements $|\Delta E|<100\mev$ and $\mes>5.2\gevcc$. While the dominant background
   comes from continuum events, there is still a non-negligible contribution
   from $\FourS\to\BB$ events (denoted ``\BB'' in the following). We consider separately the 
   $B \to D\pi$ background, since it differs from the signal only in the $\Delta E$ distribution. 
 For this decay mode the opposite-sign $B^-\to \Dzb \pi^-$ amplitude is suppressed by a factor $\approx
r_B\lambda^2$, where $\lambda\approx 0.22$ is the sine of the Cabibbo
angle. Therefore we expect to find a non-negligible $B\to D\pi$
background only in the same-sign sample.  

   \begin{figure*}[htb]
      \includegraphics[width=0.329\linewidth]{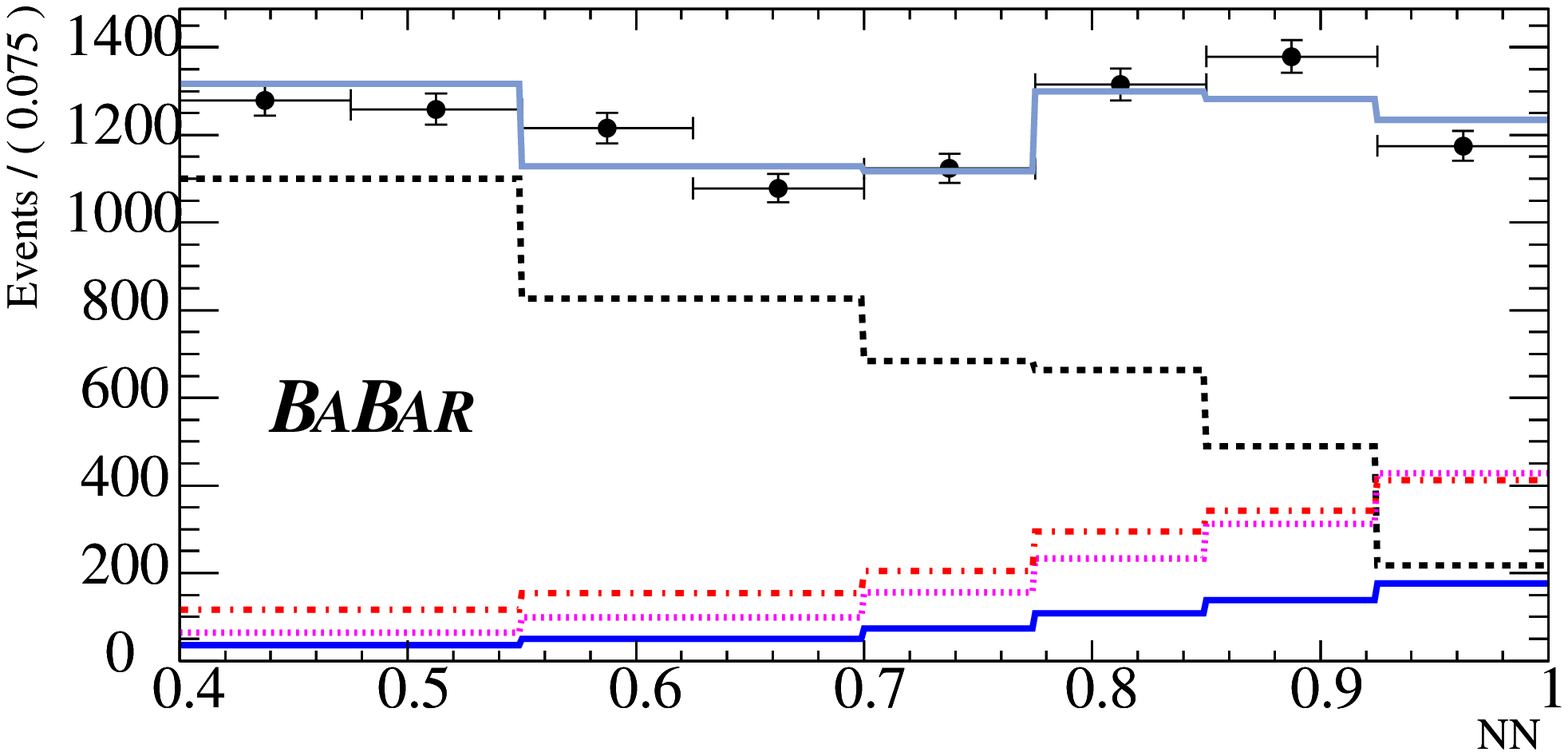}
      \includegraphics[width=0.329\linewidth]{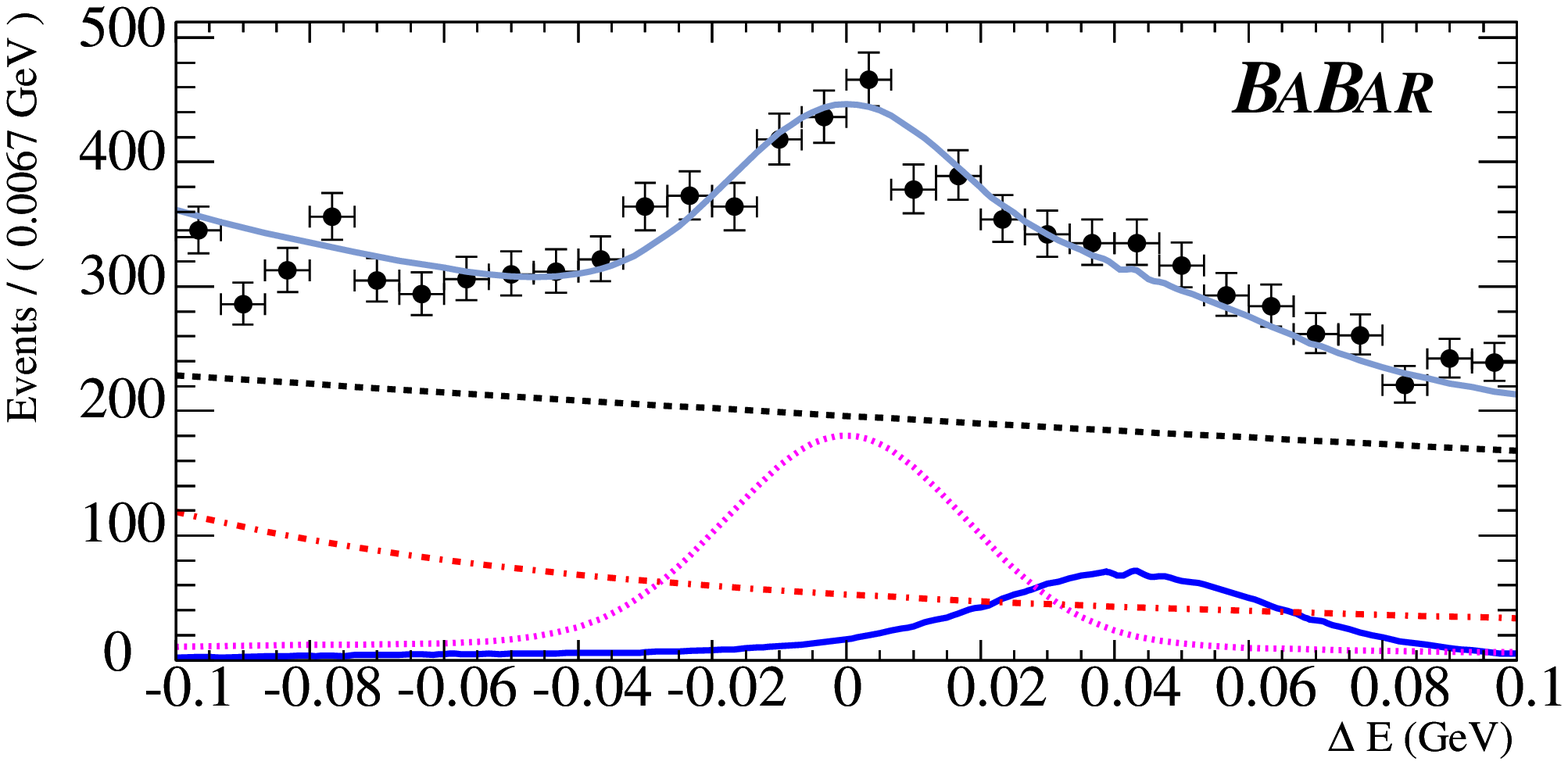}
      \includegraphics[width=0.329\linewidth]{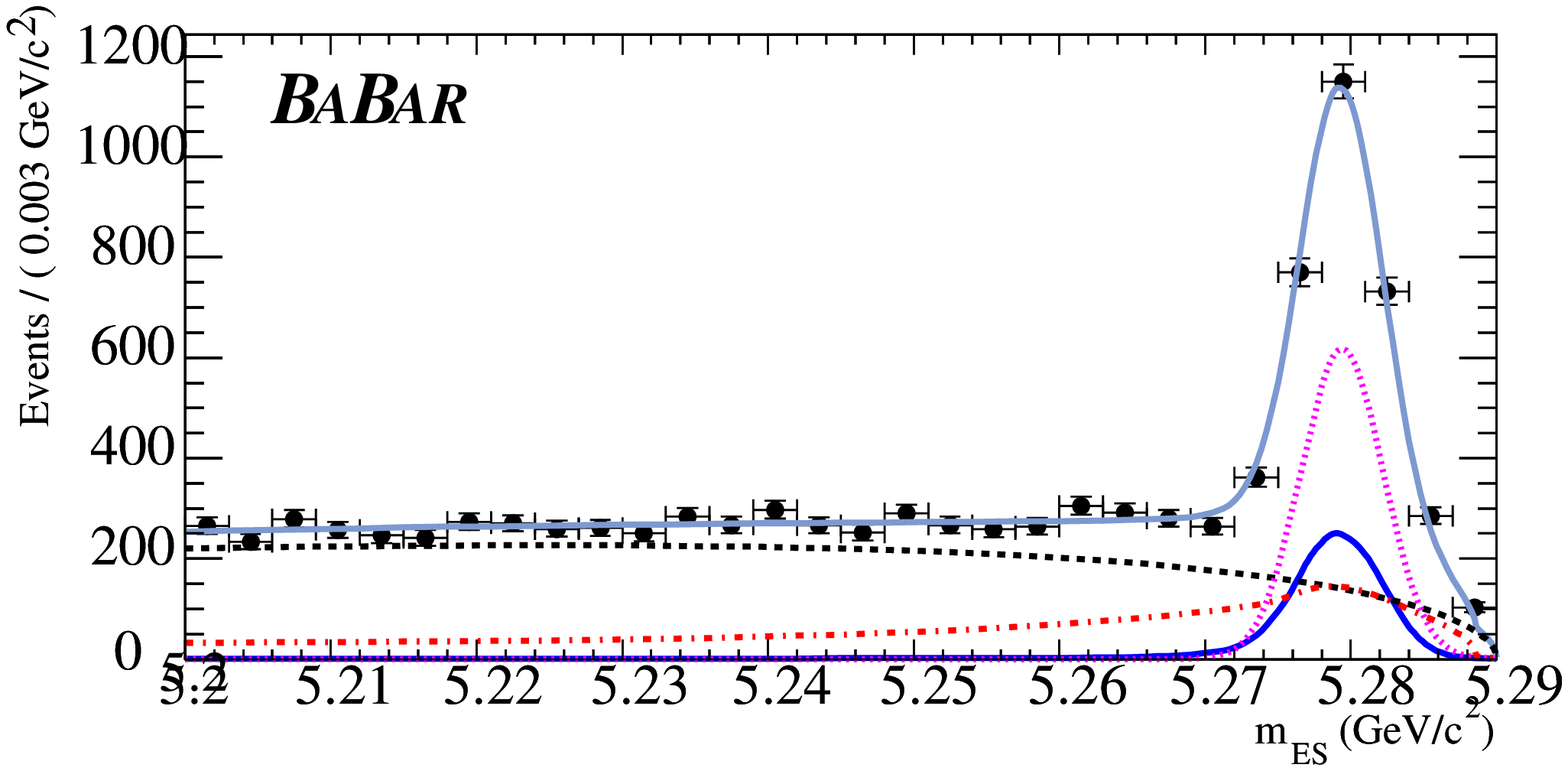} \\
      \includegraphics[width=0.329\linewidth]{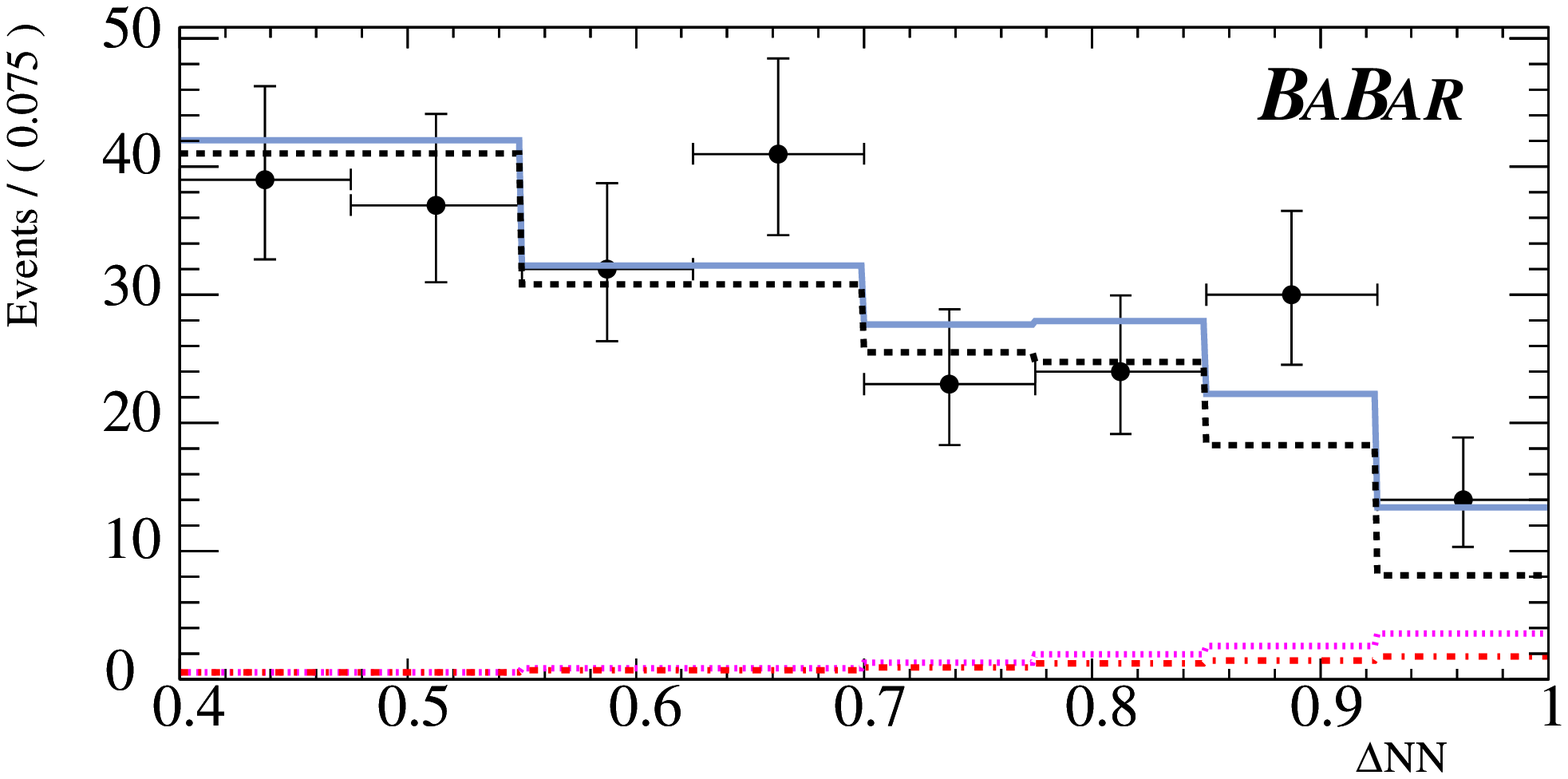}
      \includegraphics[width=0.329\linewidth]{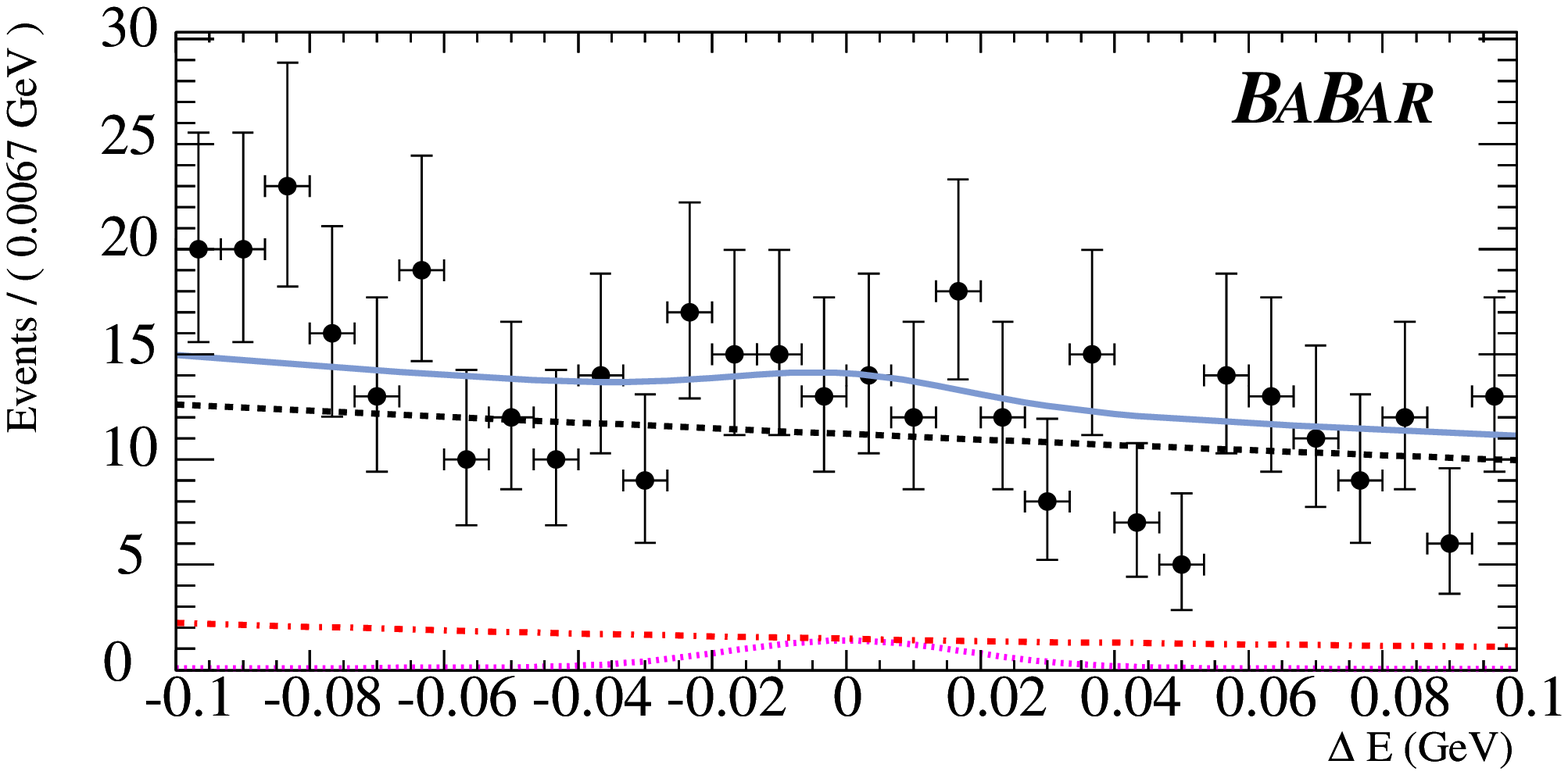}
      \includegraphics[width=0.329\linewidth]{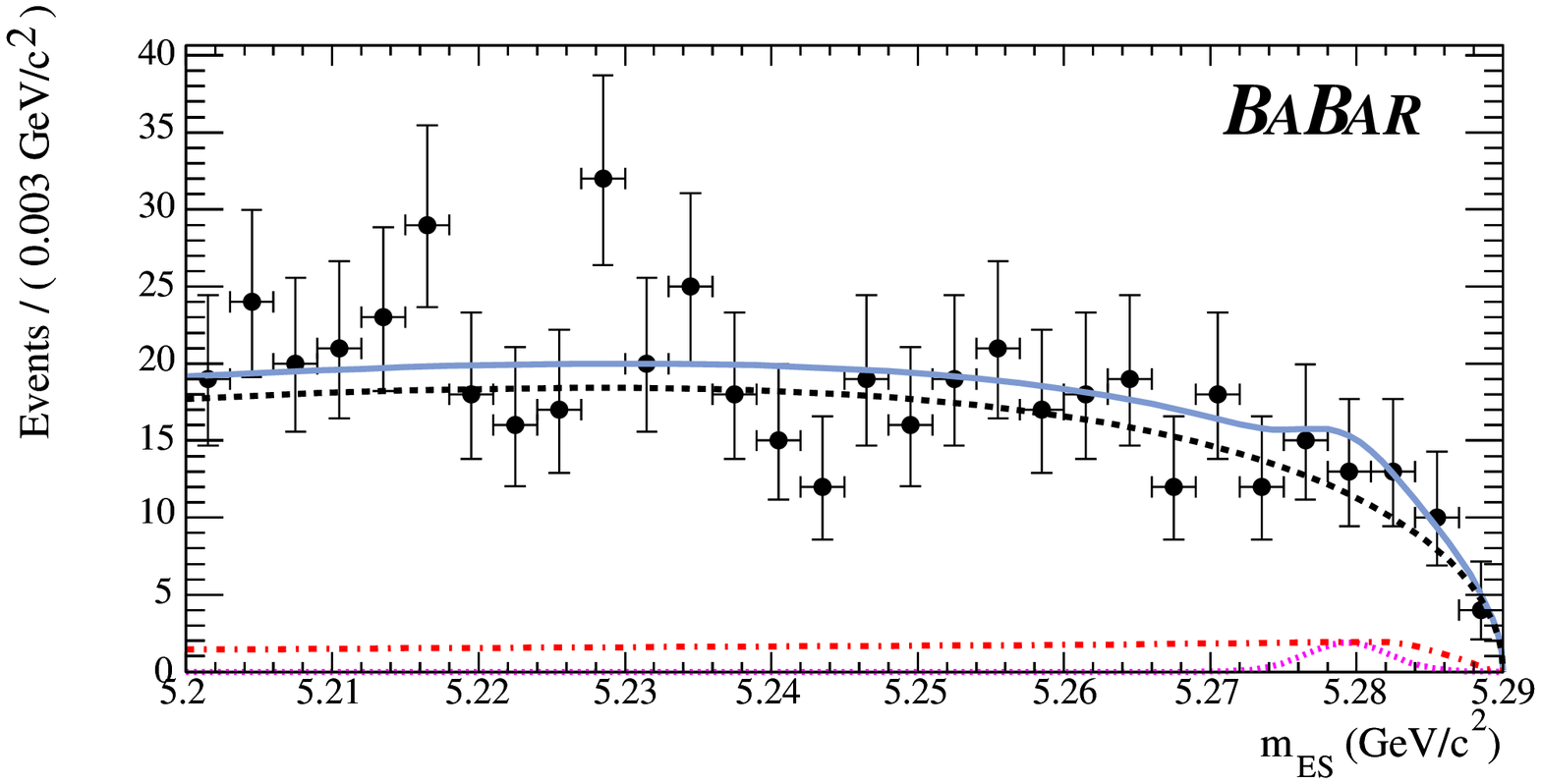} 
   \caption{Likelihood fit projections of the {\it NNet}, $\Delta E$, and $\mes$ distributions separately for the same (top) and opposite (bottom) 
sign samples. To visually enhance the signal, the distributions for the latter sample are shown
after cuts, with a 67\% signal efficiency, on the ratios between the signal and the total likelihood of all the 
 variables other than the one shown.
  The points with error bars represent the data, while
the dashed,  dash-dotted, and solid  lines represent the contributions from continuum, $\BB$, and $D\pi$ backgrounds, respectively.
The dotted line represents the signal contribution, visible only in the same-sign sample. } 
   \label{fig:fit}
   \end{figure*}
\section {Likelihood Fit and Results}

   The signal and background yields are extracted by maximizing the extended likelihood
   ${\cal {L}}= e^{-N^\prime}\prod_{i=1}^N{\cal{L}}_i(\vec{x_i})/N!$. 
Here $N^\prime=N_{DK}+N_{cont}+N_{BB}+N_{D\pi}$ 
   is the sum of the yields of the signal and the three background contributions (including both the 
same-sign and the opposite-sign components),
   $\ix=\{NNet,\Delta E,\mes\}$, and the likelihood of the individual events (${\cal{L}}_i$) is defined as
   \begin{eqnarray}
\label{eq:pdf}
      {\cal{L}}_i(\vec{x_i})=\frac{N_{DK}}{1+\rads}f^{RS}_{DK}(\vec{x_i})
      &+&\frac{N_{cont}}{1+R_{cont}}f_{cont}^{RS}(\vec{x_i}) \nonumber \\
      +\frac{N_{BB}}{1+R_{BB}}f^{RS}_{BB}(\vec{x_i})
      &+&N_{D\pi}f_{D\pi}(\vec{x_i})
   \end{eqnarray}
for same-sign events and
\begin{eqnarray}
	{\cal{L}}_i(\vec{x_i})=\frac{N_{DK}\rads}{1+R_{ADS}}f^{WS}_{DK}(\vec{x_i})
	&+&\frac{N_{cont}R_{cont}}{1+R_{cont}}f^{WS}_{cont}(\vec{x_i}) \nonumber \\ 
	+\frac{N_{BB}R_{BB}}{1+R_{BB}}f^{WS}_{BB}(\vec{x_i})&& 
   \end{eqnarray}
for opposite-sign events. In these equations
we have defined $R$ parameters for the backgrounds analogous to those 
for the signal, defined in Eq.~\ref{eq:rads}. 
The individual probability densitity functions (PDFs) $f$ are derived
from MC and are built as the product of one-dimensional
distributions of the three variables. The only exception is the \mes
and $\Delta E$ PDF for the $D\pi$ background, where we use a two-dimensional non-parametric 
distribution~\cite{KEYS} due to a non-negligible correlation between 
these two variables. The {\it NNet} distributions are all modeled with
a histogram with eight bins between 0.4 and 1. The $\mes$ distributions are modeled with a 
Gaussian in the case of the signal, a
threshold function~\cite{argus} in the case of the continuum background, and the sum of a threshold
function and a Gaussian function with an exponential tail in the case of the $\BB$ background. 
Finally, the $\Delta E$ distributions are parametrized with the sum of two Gaussians in the case of 
the signal, an exponential in the case of the continuum background, and a 
sum of two exponentials in the case of the $\BB$ background.  
For $\mes$ and $\Delta E$ of the $\BB$ and continuum background, we 
use different parameters for same-sign and opposite-sign sample.

We perform the fit by floating the four total yields ($N_{DK}$, $N_{cont}$, $N_{BB}$, and $N_{D\pi}$), the three $R$ variables 
and the shape parameters of the threshold function used to parametrize the $m_{ES}$ distribution for the same- and opposite-sign continuum 
background separately.
Figure~\ref{fig:fit} shows the distributions of the three variables in the selected sample, with the likelihood projections overlaid. 
The fit yields $\rads={\radsval}^{ \radsstatp}_{\radsstatm} $, $N_{DK}=(14.7\pm0.6)\times10^2$, $N_{cont}=(239.3\pm2.1)\times10^2$,
$N_{BB}=(25.5\pm1.6)\times 10^2$, $N_{D\pi}=(6.7\pm0.4)\times10^2$, $R_{cont}=3.05\pm0.07$, $R_{BB}=0.42\pm0.07$.

Eq. \ref{eq:pdf} assumes equal efficiencies for the same- and 
opposite-sign signal samples, regardless of the difference in the Dalitz structure. This has been 
demonstrated to be true in MC within a relative statistical error of 
4\%. We then consider this as a systematic error on \rads.
We also repeat the fit by varying the PDF parameters obtained from MC within their
statistical errors and by estimating $f_{cont}^{RS/WS}$ on off-resonance data and $f_{DK}^{RS/WS}$
on exclusively reconstructed $D\pi$ events. To account for the observed variations, we assign a 0.0076
systematic error on \rads. The uncertainty due to $B$ decays with distributions similar to the signal, in 
particular $B\to D^{(*)}\pi$, $D^*K$, $D^{(*)}K^*$, and $KK\pi\piz$,
is estimated by varying their  branching fractions within their known errors and found to be $6\cdot 10^{-5}$ on \rads, and therefore negligible.
The quality of the simulation of $B$ decays to final states with charm 
mesons that might mimic the signal has been checked by
comparing data and MC samples in the sidebands of the $\Delta E$ 
distribution where these decays dominate. Similarly, we searched the 
sidebands of the $m_{D^0}-m_{\pi^0}$ distribution for background 
from charmless $B$ decays and found no evidence of it.

Following a Bayesian approach, we extract $r_B$ by defining the posterior distribution
\begin{equation}
{\cal {L}}(r_B)=\frac{\int p(r_{B},r_D,\xi)
{\cal {L}}(\rads(r_B,r_D,\xi)) d r_D d\xi}{
\int p(r_{B},r_D,\xi){\cal {L}}(\rads(r_{B},r_D,\xi)) dr_D
d\xi dr_B},
\end{equation}
where $\xi=C\cos\gamma$,  $\rads(r_{B},r_D,\xi)$ is given in Eq.~\ref{eq:rads}, and $p(r_B,r_D,\xi)$ 
is the prior distribution for these
three quantities. They are considered uncorrelated, with $\xi$
 and $r_B$  uniformly distributed  
in the range of $[-1,1]$ and $[0,1]$ respectively.
 The prior distribution for $r_{D}$ is a Gaussian
consistent with $r^2_D=(0.214\pm0.008\pm0.008)\%$~\cite{kpipiz}.
 The likelihood ${\cal {L}}(\rads)$ is obtained 
by convolving the  likelihood returned by the fit with a Gaussian of width \radssys, equivalent to the systematic uncertainty. 

\begin{figure}[htb]
     \includegraphics[width=0.445\linewidth]{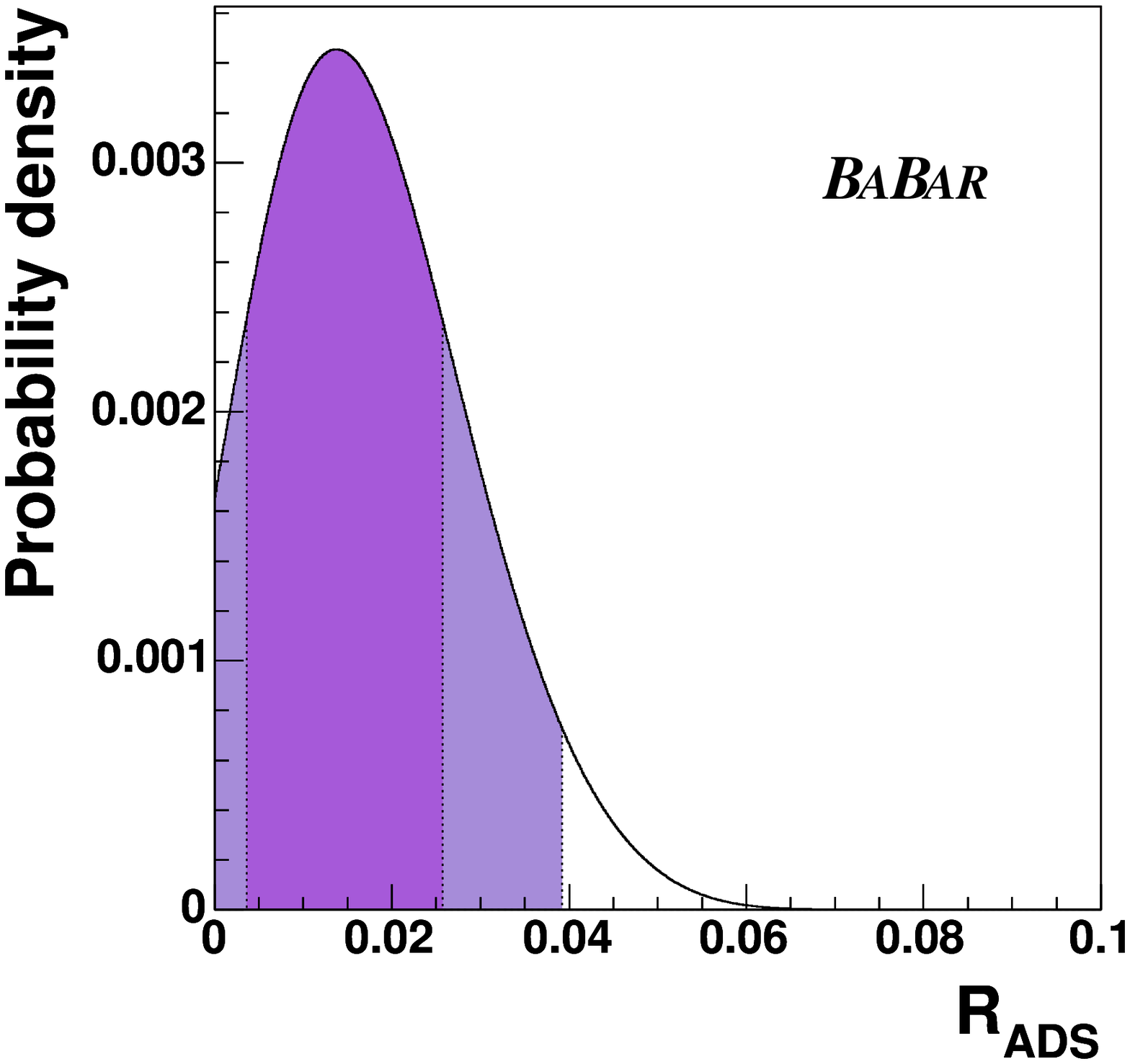}
     \includegraphics[width=0.445\linewidth]{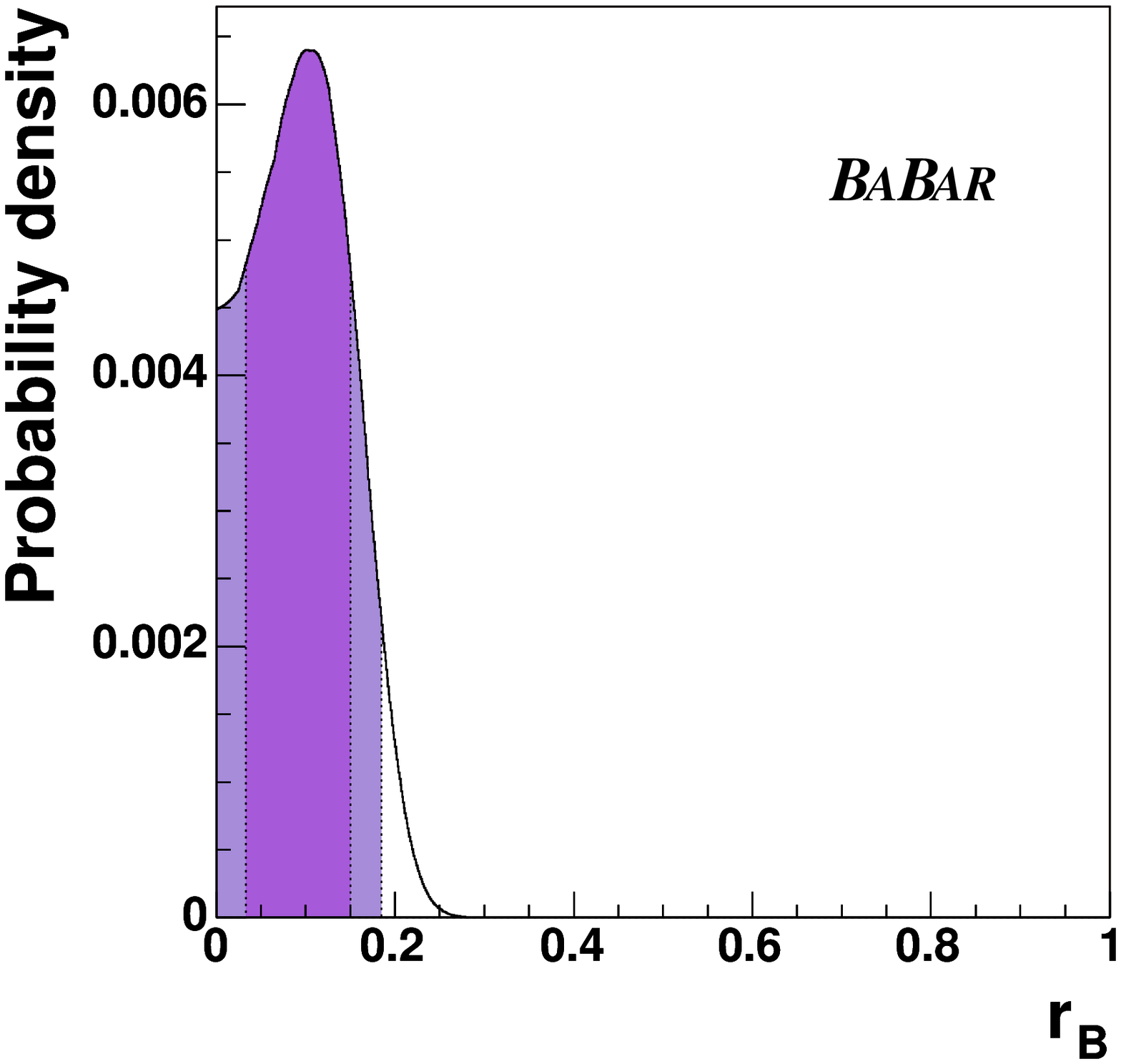}
     \caption{Likelihood function for $R_{ADS}$ (left) and $r_B$ (right).
      The latter is obtained in a bayesian approach, assuming flat prior distributions for 
       $r_{B}$ and $\xi=C \cos\gamma$.
       The 68\% and 95\% regions are shown in dark and light shading respectively.}
   \label{fig:likelihood_curve}
   \end{figure}

Figure~\ref{fig:likelihood_curve} shows ${\cal {L}}(R_{ADS})$ and ${\cal {L}}(r_B)$. 
We set a 95\% confidence-level (C.L.) limit by integrating  the 
likelihood, starting from $R_{ADS}=0$ ($r_B=0$), thus excluding unphysical
values, and we define the
68\% C.L. region, for each variable $r=R_{ADS}$ or $r_B$, as the interval where ${\cal {L}}(r)>{\cal {L}}_{min}$ and $68\%=\int_{{\cal {L}}(r)>{\cal {L}}_{min}}
{\cal {L}}(r) dr$. 

\section{Conclusions}

   In summary, we measure the ratio 
   of the rate for the $B^{\pm} \to [K^{\mp}\pi^{\pm}\piz]_D K^{\pm}$  
decay to the favored decay
   $B^{\pm} \to [K^{\pm}\pi^{\mp}\piz]_D K^{\pm}$ to be
  $\rads =\radsmeas$. While this result is consistent with and similar in sensitivity to the completely independent 
previously published results~\cite{oldADS}, it is obtained using a different $D$ decay mode.
 Because the measurement is not statistically significant, we set a
   95\% C.L. limit  $\rads < \radslim$. 
   We use this information to infer the ratio between the rates
   of the $B^- \to \Dzb K^-$ and
   $B^- \to D^0 K^-$ decays to be $r_B =\rbmeas$ 
   and consequently set a limit $r_B<\rblim$ at 95\% C.L.

We are grateful for the excellent luminosity and machine conditions
provided by our \pep2\ colleagues, 
and for the substantial dedicated effort from
the computing organizations that support \babar.
The collaborating institutions wish to thank 
SLAC for its support and kind hospitality. 
This work is supported by
DOE
and NSF (USA),
NSERC (Canada),
IHEP (China),
CEA and
CNRS-IN2P3
(France),
BMBF and DFG
(Germany),
INFN (Italy),
FOM (The Netherlands),
NFR (Norway),
MIST (Russia),
MEC (Spain), and
PPARC (United Kingdom). 
Individuals have received support from the
Marie Curie EIF (European Union) and
the A.~P.~Sloan Foundation.


\begin{thebibliography}{99}

   \bibitem{cpv} BaBar Collaboration, B.~Aubert {\em et al.},
   Phys. Rev. Lett. {\bf 89}, 201802 (2002); Belle Collaboration,
   K. Abe {\em et al.}, Phys. Rev. {\bf D66}, 071102 (2002).



   \bibitem{dk1} M.~Gronau and D.~Wyler, Phys. Lett. {\bf B265}, 172 
(1991);
   M.~Gronau and D.~London, Phys. Lett. {\bf B253}, 483 (1991).

   \bibitem{dk2} D.~Atwood, I.~Dunietz, and A.~Soni, Phys. Rev. Lett. 
{\bf 78},
   3257 (1997); Phys. Rev. {\bf D63}, 036005 (2001).

   \bibitem{cc}Charge conjugation is implied throughout the paper. 
Also, we use 
    the notation $B^- \to [h_1^+h_2^-\piz]_D h_3^-$ 
   (with $h_i=\pi$ or $K$) for the decay chains $B^- \to \dbarp h_3^-$, 
where $\dbarp$ is either
   a \Dz or a \Dzb and $\dbarp~\to h_1^+ h_2^-\piz$.  
   We also refer to $h_3$ as the $\pi$ or $K$ from the $B$.


\bibitem{Dmix}
   Y.~Grossman, A.~Soffer and J.~Zupan,
 Phys.\ Rev. {\bf D72} (2005) 031501
   \bibitem{dalitz}
BaBar Collaboration, B.~Aubert {\it et al.} ,
  Phys.\ Rev.\ Lett.\  {\bf 95}, 121802 (2005).
   \bibitem{oldADS}
BaBar Collaboration, B.~Aubert {\it et al.},
  Phys.\ Rev. {\bf D72}, 032004 (2005);
Belle Collaboration, M.~Saigo {\it et al.},
  Phys.\ Rev.\ Lett.\  {\bf 94} 091601 (2005).

   \bibitem{kpipiz}
BaBar Collaboration, B.~Aubert {\it et al.}, 
Phys.\ Rev.\ Lett.\
{\bf 97}, 221803 (2006).   
\bibitem{rdkpi}
BaBar Collaboration, B.~Aubert {\it et al.},
  Phys.\ Rev.\ Lett.\  {\bf 91}, 171801 (2003).
   \bibitem{pep2} PEP-II Conceptual Design Report, SLAC-0418 (1993).
   \bibitem{babar}
BaBar Collaboration, B.\ Aubert {\em et al.},
   Nucl. Instrum. and Methods Phys. Res., Sec. {\bf A 479}, 1 (2002).
   \bibitem{PDG} 
 Particle Data Group, S.~Eidelman {\it et al.},
  Phys.\ Lett. {\bf B592}, 031501 (2005).   

\bibitem{KEYS}
K.~Cranmer, \cpc{136}, 198 (2001).
\bibitem{argus}
ARGUS Collaboration, H.~Albrecht {\em et al.}, \zpc {\bf 48}, 543 
(1990).

   \end{thebibliography}
\end{document}